\documentclass[10pt, twocolumn]{article} 
\usepackage[main=english,french,spanish]{babel} 
\usepackage{fontspec} 

\usepackage[a4paper, width=170mm,top=25mm,bottom=30mm]{geometry}
\usepackage[indent=20pt]{parskip} 

\usepackage[
  sorting=none,
  style=numeric,
  backend=biber,
  giveninits=true
]{biblatex} 
\bibliography{references.bib}
\AtEveryBibitem{%
  \clearfield{urlyear}%
  \clearfield{urlmonth}%
  \clearfield{urlday}%
}

\usepackage[affil-it]{authblk} 

\usepackage{graphicx}
\graphicspath{{images/}{figures/}}

\usepackage{csquotes} 

\usepackage[subpreambles=true]{standalone}

\usepackage{listings} 

\usepackage{float} 
\usepackage{amsmath, amssymb, amsthm} 
\usepackage[colorlinks=true, allcolors=blue]{hyperref}  
\usepackage[noabbrev, capitalise]{cleveref} 
\AddToHook{cmd/appendix/before}{\crefalias{section}{appendix}}

\usepackage{xpatch}
\xpatchbibmacro{name:andothers}{%
  \bibstring{andothers}%
}{%
  \bibstring[\emph]{andothers}%
}{}{}

\usepackage[version=4]{mhchem} 
\usepackage[multi-part-units=single, separate-uncertainty=true, range-units=single]{siunitx} 
\DeclareSIUnit\Td{Td}
\DeclareSIUnit\Gy{Gy}
\DeclareSIUnit\torr{Torr}
\usepackage{physics} 
\AtBeginDocument{\RenewCommandCopy\qty\SI}
\ExplSyntaxOn
\msg_redirect_name:nnn { siunitx } { physics-pkg } { none }
\ExplSyntaxOff

\usepackage[dvipsnames]{xcolor} 

\usepackage{todonotes} 
\setuptodonotes{inline, size=\small}

\usepackage{caption}
\usepackage{subcaption}

\renewcommand{\dd}{\mathop{}\!\mathrm{d}}

\begin{document}

\twocolumn[
\begin{center}

{\LARGE\bfseries
Charge Collection Efficiency in Air-Vented Plane-Parallel Ionisation Chambers at Ultra-High Dose Rates
\par}

\vspace{0.5em}

{\Large
A Self-Consistent Garfield++ Monte Carlo Model Including Space-Charge Effects and Ion Recombination
\par}

\vspace{1em}

{\large
Pierre Gérard Ortega$^{1,*}$,
Gilles De Lentdecker$^{1}$,
Quentin Flandroy$^{2}$,
Jarrick Nys$^{2}$
\par}

\vspace{0.5em}

{\small
$^{1}$ Service de Physique des particules élémentaires (IIHE), Université libre de Bruxelles (ULB), Brussels, Belgium\\
$^{2}$ Ion Beam Applications SA, Louvain-la-Neuve, Belgium
\par}

\vspace{0.5em}

{\small \today\par}

\vspace{1.5em}

\begin{minipage}{0.85\textwidth}
{\centering\small\bfseries Abstract\par}

\vspace{0.5em}

\small
Ultra-high dose rate (UHDR) irradiation used in FLASH radiotherapy can lead to significant reductions in charge collection efficiency (CCE) in plane-parallel ionisation chambers (PPICs) due to enhanced ion recombination, while the high charge densities involved can also induce strong space-charge-induced electric-field distortions. To investigate these coupled effects, we extended the Garfield++ Monte Carlo particle-transport framework by implementing ion--ion recombination and self-consistent space-charge electric field calculations.

The developed Monte Carlo model couples particle transport, electron attachment, recombination processes, and dynamic electric-field distortions. The implementation was validated against analytical and numerical models from the literature. Excellent agreement was obtained for the free electron fraction (FEF), CCE, induced current, and electric field evolution under UHDR conditions.

The simulations show that space charge can locally increase the electric field by more than a factor of four or reduce it to nearly zero in some regions of the chamber. The results further suggest that the reduction of the CCE under UHDR conditions is mainly driven by the decrease of the FEF caused by electric-field-dependent electron attachment. This finding indicates that the complex problem of recombination under UHDR conditions may be largely governed by the evolution of the FEF, opening promising perspectives for the development of improved analytical models and real-time correction methods for ionisation chamber dosimetry under UHDR irradiation conditions.

This work provides a flexible and self-consistent Monte Carlo framework for investigating recombination phenomena and improving dosimetry models for FLASH radiotherapy.

\par\vskip\baselineskip\noindent


\end{minipage}

\vspace{2em}

\end{center}
]
\renewcommand{\thefootnote}{\fnsymbol{footnote}}
\footnotetext[1]{Corresponding author:
\href{mailto:pierre.gerard.ortega@ulb.be}{pierre.gerard.ortega@ulb.be}}
\renewcommand{\thefootnote}{\arabic{footnote}}

\section{Introduction}
\label{sec:Introduction}
Since the discovery of what is now referred to as the FLASH effect by~\textcite{favaudon_2014} in 2014, research on ultra-high dose rate (UHDR) radiation ($\geq \SI{40}{\gray\per\second}$) has grown rapidly. The reduced toxicity observed in normal tissues during UHDR irradiation, combined with equivalent or even enhanced tumour control, offers promising perspectives for radiotherapy. Subsequent studies have confirmed normal-tissue sparing in a variety of preclinical models and tissues. In particular, FLASH irradiation was shown to improve survival following abdominal irradiation in mice~\cite{loo:2017}, preserve cognitive function after whole-brain irradiation~\cite{montay-gruel:2017}, and reduce brain toxicity using both electron and X-ray beams~\cite{montay-gruel:2018,montay-gruel:2019}. The FLASH effect has also been demonstrated in larger animals and veterinary cancer patients~\cite{vozenin:2019}, further highlighting its potential to widen the therapeutic window of radiotherapy.

Ionisation chambers (ICs) are the reference dosimeters recommended by the International Atomic Energy Agency (IAEA)~\cite{internationalatomicenergyagency_2024} for the determination of absorbed dose to water. In particular, plane-parallel ionisation chambers (PPICs) are widely used in radiotherapy both for reference dosimetry and for beam monitoring during patient treatment. However, dosimetry under UHDR conditions remains challenging~\cite{subiel:2024}, and experimental studies of commercially available PPICs have shown a strong dependence of their charge collection efficiency on dose per pulse, while also highlighting the need for improved ion recombination models for accurate dosimetry under UHDR conditions~\cite{bourgouin:2023}.

To accurately determine the absorbed dose, several correction factors must be applied during IC calibration. One of these factors, $k_s$, accounts for incomplete charge collection caused by ion recombination.

Under conventional radiotherapy conditions, ion recombination corrections are commonly performed using the two-voltage method, which is based on the model originally developed by \textcite{boag_1950} in \citeyear{boag_1950}. This model was later extended by \textcite{boag_1996} in \citeyear{boag_1996} to account for the fraction of electrons escaping attachment to electronegative molecules, referred to as the free electron fraction (FEF). This contribution is not considered in the two-voltage methodology recommended by TRS-398~\cite{internationalatomicenergyagency_2024} and TG-51~\cite{almond:1999}. However, it is now well established~\cite{laitano_2006, fenwick_2023, orts_2024, gerardortega_2024} that these analytical models are no longer valid under the UHDR conditions required for FLASH radiotherapy. In particular, electric-field distortions induced by space charge can no longer be neglected. Several studies~\cite{kranzer_2021, baack_2022, boissonnat_2015} have shown that these distortions can locally exceed the nominal applied electric field by several times.

To better understand recombination effects in PPICs under UHDR pulsed beams, we extended the Garfield++~\cite{garfieldpp} framework and developed a self-consistent Monte Carlo simulation coupling particle transport, electron attachment, ion--ion recombination, and space-charge effects. The flexibility of Garfield++ enables simulations for arbitrary gas mixtures, atmospheric conditions, IC geometries, incident particle types, and beam spatial and temporal structures, allowing comprehensive investigations of recombination phenomena under a wide range of irradiation conditions.

\section{Physical background and literature review}
This section reviews the physical mechanisms governing charge transport and recombination in air-filled ionisation chambers. Particular emphasis is placed on electron attachment, ion transport, and ion--ion recombination processes.

\subsection{Electron swarm parameters and attachment processes}
\label{sec:Electron_swarm_parameters}
The dominant mechanism responsible for the formation of negative ions in air is the attachment of electrons to oxygen molecules. Unlike noble gases, often used in gaseous detectors to avoid such phenomena, oxygen requires only a relatively low collision energy to capture free electrons. Two different attachment mechanisms exist: two-body and three-body attachment.

Two-body attachment occurs through a dissociative process involving the oxygen molecule and requires a minimum dissociation energy threshold of \SI{5.61}{\eV}~\cite{deurquijo_2024}:
\begin{equation}
	\ce{e- + O2 -> O- + O}.
\end{equation}

Water molecules present in humid air can also undergo dissociation and form negative ions following electron attachment, although their dissociation energy is higher, around \SI{6.32}{\eV}:
\begin{equation}
	\ce{e- + H2O -> OH- / O- / H-}.
\end{equation}

At low electron energies, three-body attachment becomes dominant and proceeds through a two-step process. In this mechanism, a third body, generally \ce{O2} or \ce{H2O}, is required to de-excite the metastable oxygen anion~\cite{deurquijo_2024, blum_2008, bischneider_1982}:
\begin{subequations}
\begin{align}
	\ce{e- + O2 -> &O2^{-*}} \\
	\ce{&O2^{-*} + M -> O2- + M + E_{kin}}.
\end{align}
\end{subequations}

\begin{figure}[t]
	\centering
    \includegraphics[width=1\linewidth]{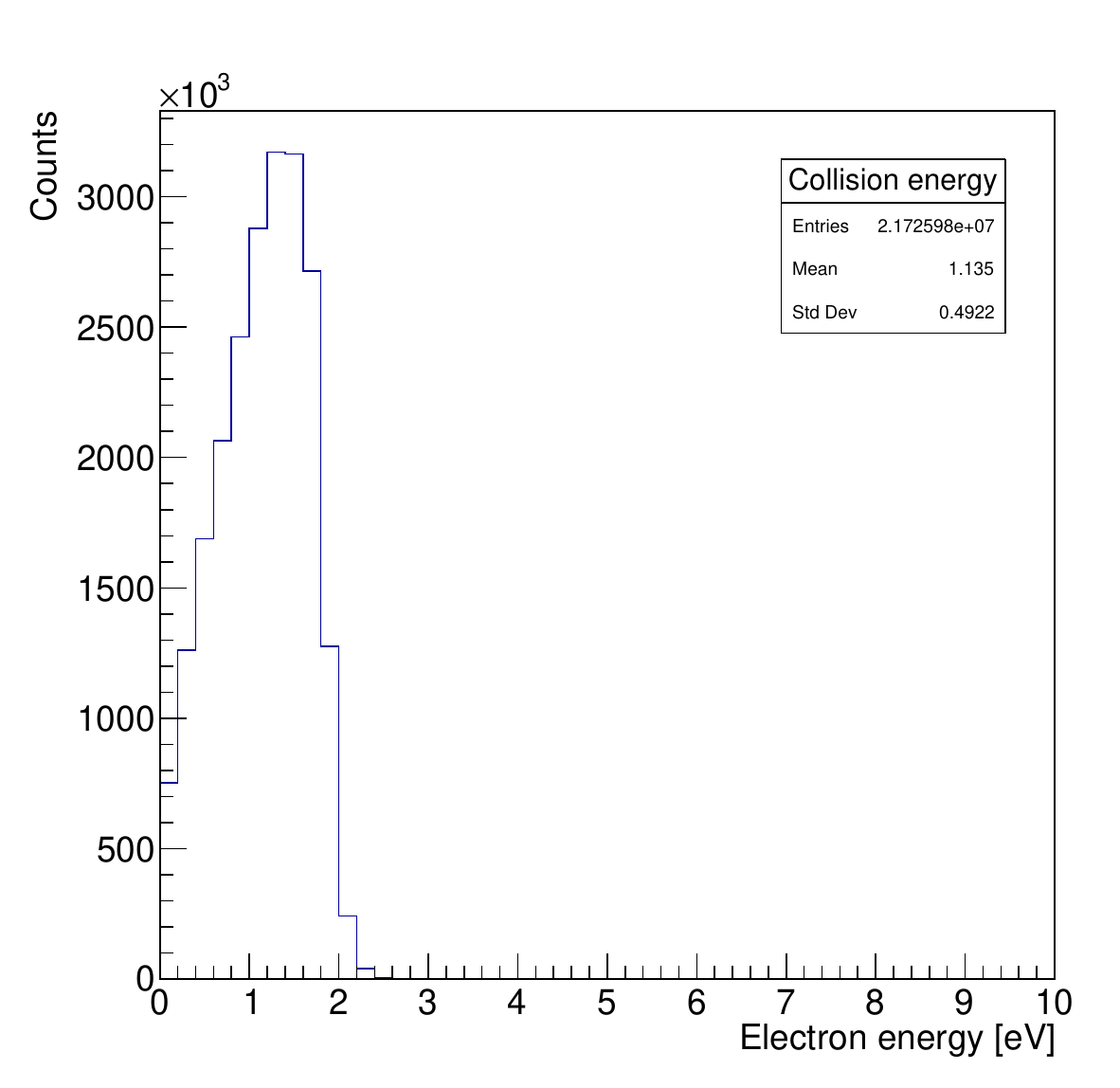}
	\caption{Electron collision energy at \SI{20}{\Td} in humid air with \SI{2}{\percent} \ce{H2O}.}
	\label{fig:electron_energies}
\end{figure}

Using Magboltz~\cite{magboltz}, we simulated electron transport in air under a reduced electric field of \SI{20}{\Td} ($\SI{1}{\Td}=\SI{e-21}{\V\m\squared}$), corresponding to an electric field of approximately \SI{5.0}{\kilo\V\per\cm} at standard temperature and pressure (STP). The simulation yielded a mean electron collision energy of \SI{1.14}{\eV} (\cref{fig:electron_energies}), indicating that only the three-body attachment process is energetically accessible under these conditions. This result is consistent with the findings reported by \textcite{deurquijo_2024}.

\begin{figure*}[t]
\centering
\begin{subfigure}[b]{0.48\textwidth}
\includegraphics[width=\textwidth]{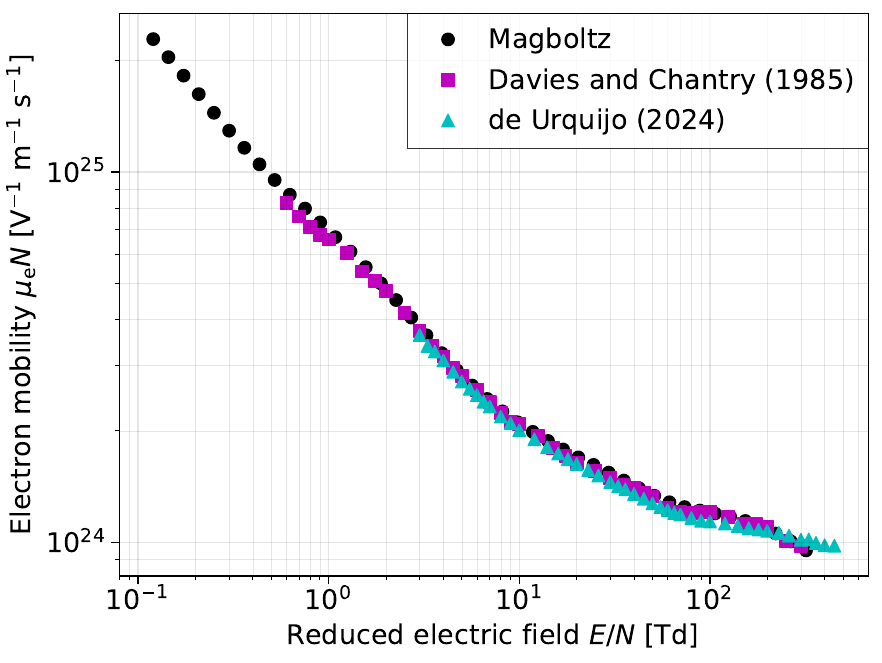}
\caption{Electron mobility in dry air.}
\end{subfigure}
\hfill
\centering
\begin{subfigure}[b]{0.48\textwidth}
\includegraphics[width=\textwidth]{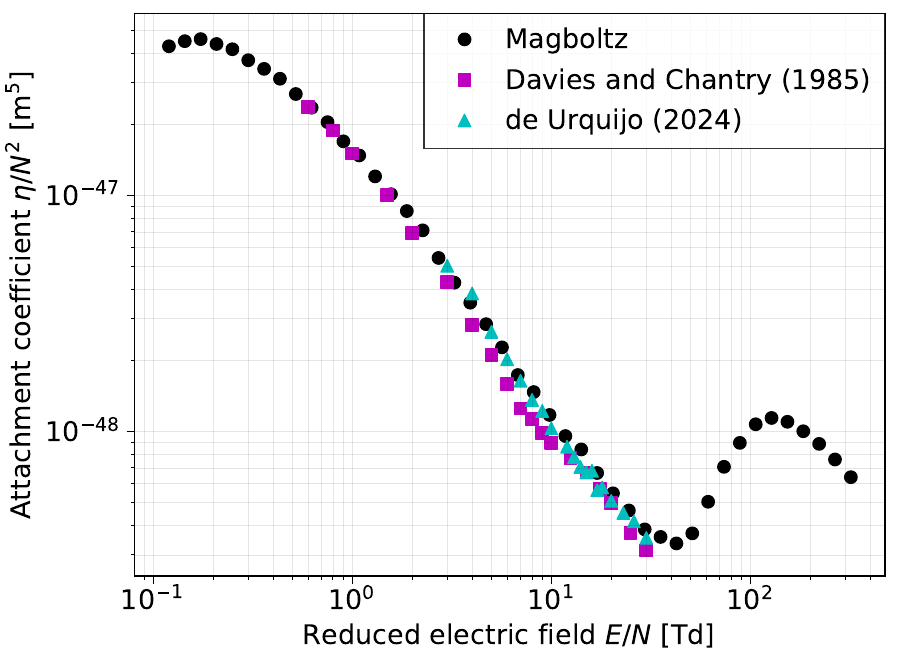}
\caption{Electron attachment coefficient in dry air.}
\end{subfigure}
\begin{subfigure}[b]{0.48\textwidth}
\includegraphics[width=\textwidth]{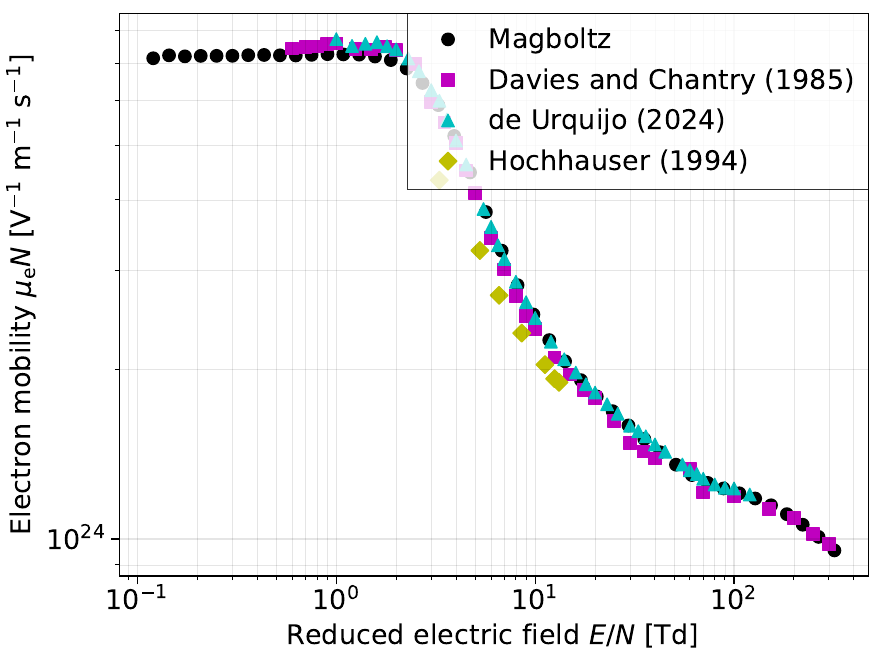}
\caption{Electron mobility in humid air (\SI{1.2}{\percent} \ce{H2O} and \SI{2}{\percent} \ce{H2O}).}
\end{subfigure}
\hfill
\begin{subfigure}[b]{0.48\textwidth}
\includegraphics[width=\textwidth]{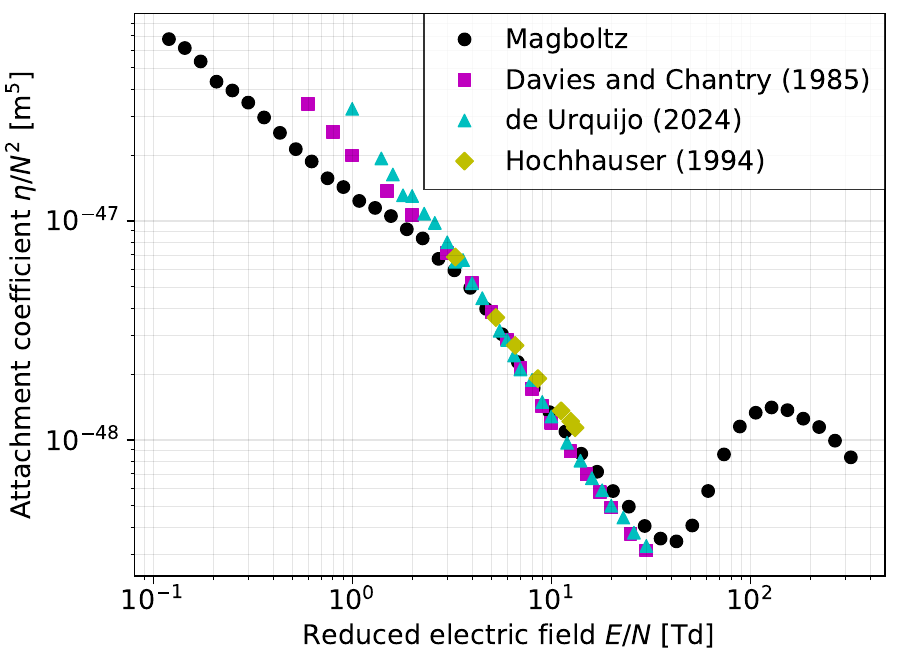}
\caption{Electron attachment coefficient in humid air (\SI{1.2}{\percent} \ce{H2O} and \SI{2}{\percent} \ce{H2O}).}
\end{subfigure}
\caption{Comparison of the electron attachment coefficient and electron mobility in dry and humid air obtained from Magboltz simulations with experimental measurements reported in the literature. Experimental data correspond to dry air and humid air with \SI{2}{\percent} \ce{H2O}~\cite{deurquijo_2024, davies_1985}, while data from \textcite{hochhaeuser_1994} correspond to humid air with \SI{1.2}{\percent} \ce{H2O}.}
\label{fig:attachment_and_mobility}
\end{figure*}

To validate the transport parameters obtained with Magboltz, we simulated electron transport in dry air and humid air containing \SI{2}{\percent} \ce{H2O}. Simulations were performed at \SI{293.15}{\kelvin} and \SI{1013.25}{\hecto\pascal}, using a gas composition of \SI{76.54}{\percent} \ce{N2}, \SI{20.54}{\percent} \ce{O2}, \SI{0.92}{\percent} \ce{Ar} and \SI{2}{\percent} \ce{H2O} for humid air. The resulting electron mobility $\mu_e$ and electron attachment coefficient $\eta$ were compared with experimental measurements reported in the literature~(\cref{fig:attachment_and_mobility}). The comparison includes measurements from \textcite{deurquijo_2024} and \textcite{davies_1985} for dry air and humid air containing \SI{2}{\percent} \ce{H2O}, as well as measurements from \textcite{hochhaeuser_1994} for humid air containing \SI{1.2}{\percent} \ce{H2O}. This difference in water concentration is expected to affect the electron mobility, particularly at low electric fields. Additional Magboltz simulations show that increasing the \ce{H2O} concentration from \SI{1.2}{\percent} to \SI{2}{\percent} increases the mobility by approximately \SI{38}{\percent} at \SI{100}{\volt\per\cm}, while the increase is reduced to approximately \SI{1.6}{\percent} at \SI{6}{\kilo\volt\per\cm}.

Small discrepancies are observed in the attachment coefficient, mainly at low reduced electric fields. However, these differences occur well outside the operating range considered in this work. Typical ionisation chambers operate at electric fields around \SI{3}{\kilo\volt\per\centi\meter}, corresponding to a reduced electric field of approximately \SI{12}{\Td} in air.

\subsection{Ion swarm parameters and recombination}
\label{sec:ion_swarm_parameters}
Following the ionisation of the incident particle, the dominant positive ions formed in air are \ce{N2+} and \ce{O2+}. As discussed in \cref{sec:Electron_swarm_parameters}, free electrons rapidly attach to electronegative molecules in air, forming mainly \ce{O2-} anions. In air, the attachment time depends on pressure, temperature, humidity, and electric field strength. In PPICs, the characteristic attachment time is expressed as
\begin{equation}
    \tau = \frac{1}{\eta\,v_e},
\end{equation}
where $v_e$ denotes the electron drift velocity. Under typical operating conditions, $\tau$ ranges from approximately \SI{1}{\nano\second} to \SI{100}{\nano\second}.

Very few experimental measurements of ion mobility in air are available in the literature. Ion mobility is known to depend not only on the electric field~\cite{ellis_1976, viehland_1995, zou_2016}, but also on atmospheric conditions such as pressure, temperature, and humidity~\cite{zhang_2017, zhang_2019}. During the drift phase, hydration and clustering processes involving water molecules can lead to the formation of heavier ions~\cite{franchin_2015}. These species reduce the average mobility of both cations and anions. \Textcite{zhang_2019} reported ion mobility values ranging from \SIrange{1.5e-4}{3.3e-4}{\meter\squared\per\volt\per\second} for anions and from \SIrange{1.4e-4}{2.0e-4}{\meter\squared\per\volt\per\second} for cations.

A similarly large spread exists for the ion--ion recombination coefficient $\alpha$, with reported values mainly ranging between \SIrange[range-phrase = { and }]{0.9e-12}{1.6e-12}{\cubic\metre\per\second}~\cite{boag_1950, mcgowan_1965, gotz_2017}. Furthermore, \textcite{franchin_2015} demonstrated a strong dependence of $\alpha$ on atmospheric conditions such as temperature and relative humidity, reporting values ranging from \SIrange{0.93e-12}{1.5e-12}{\cubic\metre\per\second} for relative humidities between \SIrange[range-phrase = { and }]{0}{70}{\percent}.

Consequently, the large uncertainties associated with ion mobility and ion--ion recombination coefficients constitute one of the dominant sources of uncertainty in the physical parameters of the present simulation.

\section{Methods}
In this study, Garfield++ was used to simulate PPICs and model the physical processes occurring under UHDR irradiation conditions. To achieve this, several existing Garfield++ classes and methods were extended to account for space-charge effects and ion--ion recombination.

\subsection{Garfield++ simulation toolkit}
Garfield++~\cite{garfieldpp} is an object-oriented simulation toolkit dedicated to the study of charged particle transport and interactions in gases and semiconductors. Originally developed by Rob Veenhof and now actively maintained at CERN, Garfield++ combines Monte Carlo transport techniques with electric and magnetic field maps to simulate charged particle transport in detector systems.

The framework interfaces with several external packages, including ROOT~\cite{ROOT_NIMA_1997} for visualisation, Magboltz~\cite{magboltz} for the calculation of electron transport properties in gas mixtures, and Heed~\cite{smirnov_2005} for the simulation of primary ionisation produced by charged particles.

\begin{figure}[t]
	\centering
    \begin{tikzpicture}[scale=2, node distance = 4em, 
                      auto, label distance = 1em]

    \node (medium) [block,
                    label=left:{\parbox{10em}{
                    \footnotesize material properties 
                    \begin{itemize} 
                      \item gases \(\rightarrow\) Magboltz 
                      \item semiconductors 
                    \end{itemize}}}]  {\small Medium};
    \node (dummy)  [right of=medium,
                    label=above:{\small detector description}] {};
    \node (geo)    [block,right of = dummy]  {\small Geometry};
    \node (comp1)  [block, below of = dummy, 
                    xshift = 1em, yshift = -1em]     {};
    \node (comp2)  [block, below of = dummy,
                    xshift =.5em,yshift=-.5em]   {};
    \node (comp)   [block, below of = dummy,
                    label=left:{\parbox{8em}{
                    \footnotesize \vspace{2em} 
                    field calculation 
                    \begin{itemize} 
                      \item analytic 
                      \item field maps 
                      \item neBEM 
                    \end{itemize}}}]  {\footnotesize Component};
    \node (sensor) [block, below of = comp]  {\small Sensor};
    \node (dummy1) [below of = sensor,
                    label=below:{\small Transport}] {};
    \node (aval)   [drift, right of=dummy1,label=right:{\parbox{8em}{\footnotesize charge transport \begin{itemize} \item microscopic \item MC \item RKF \end{itemize}}}]        {\small Drift};
    \node (track)  [drift, left of=dummy1,label=left:{\parbox{8em}{\footnotesize primary ionisation \begin{itemize} \item Heed \item SRIM, TRIM \item Degrade \end{itemize}}}]      {\small Track}; 
    \path [line]   (medium) -- (comp);
    \path [line]   (geo)    -- (comp);
    \path [line]   (comp)   -- (sensor);
    \path [line,dashed]  (sensor) -- (track);
    \path [line,dashed]  (sensor) -- (aval);
  \end{tikzpicture}
	\caption{Overview of the main classes in Garfield++ and their interplay (from~\cite{schindler_2012}).}
	\label{fig:garfieldpp_structure}
\end{figure}
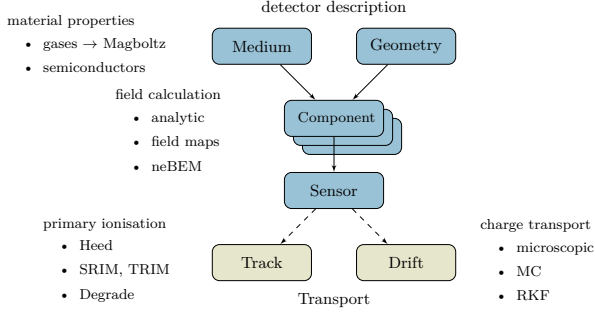

A typical Garfield++ simulation follows the structure illustrated in \cref{fig:garfieldpp_structure}. First, the detector medium (gas mixture, semiconductor, \dots) is defined using the \texttt{Medium} class. The detector geometry and associated electric or magnetic fields are then implemented using one or several \texttt{Component} classes. These components are combined through the \texttt{Sensor} class, which defines the active detector volume and manages signal calculation.

Once the detector is fully defined, dedicated transport classes can be used to simulate particle motion. The \texttt{TrackHeed} class simulates the ionisation pattern produced by relativistic charged particles. Electron transport is performed using \texttt{AvalancheMicroscopic}, which tracks electrons collision-by-collision at the microscopic level, while ion transport is handled by \texttt{AvalancheMC}, which uses Monte Carlo techniques to simulate ion drift lines.

\subsection{Implementation of recombination}
\label{sec:implementation_recombination}
In ion--ion recombination processes, a distinction is usually made between initial and volume recombination. Initial recombination occurs between ions of opposite charge created within the same ionisation track of an incident particle. It strongly depends on the linear energy transfer (LET), the medium properties, and the electric field strength. By definition, initial recombination is independent of the dose rate.

In contrast, volume recombination corresponds to recombination between ions originating from different particle tracks. This process is largely independent of the LET and becomes dominant at high dose rates. Consequently, only volume recombination was considered in the present work.

Volume recombination can be mathematically described by the following equation with the recombination coefficient $\alpha$ in \si{\cubic \m\per\s} and the positive and negative ion densities $n_{+/-}$ in \si{\per\cubic\m}.
\begin{equation}
    \frac{\dd n_{+}}{\dd t} = \frac{\dd n_{-}}{\dd t} = -\alpha \, n_{+} \, n_{-}
\end{equation}
Following this equation, the ion density decays exponentially in time, with a decay rate proportional to the density of ions of opposite charge:
\begin{equation}
    n_{+/-}(t) = n_{+/-}(t_0)\, \exp(-\alpha \, n_{-/+}\, (t-t_0)).
\end{equation}

The recombination probability over the time interval $\Delta t = t_1 - t_0$ can then be written as a Poissonian process.
\begin{subequations}
\begin{align}
    P_{rec}(+/-) &= \frac{n_{+/-}(t_0) - n_{+/-}(t_1)}{n_{+/-}(t_0)} \\
                     &= 1 - \exp(-\alpha \, n_{-/+} \, \Delta t)
    \label{eq:recombination_probability}
\end{align}
\end{subequations}

Within Garfield++, ion--ion recombination was implemented in the \texttt{AvalancheMC} class using an approach analogous to electron attachment. Inside the \texttt{DriftLine} method, called at each Monte Carlo time step to compute ion drift trajectories, the recombination probability is evaluated using~\cref{eq:recombination_probability}. A random number $p \in [0,1)$ is then generated and, if $p < P_{\mathrm{rec}}$, the ion is considered recombined and removed from the simulation. The recombination coefficient $\alpha$ can be specified through the newly implemented method \texttt{EnableRecombination(bool enable, double alpha)}.

To evaluate positive and negative ion densities, the \texttt{ComponentGrid} class was extended to compute particle density maps. This class defines a three-dimensional Cartesian domain $\Omega = [x_0, x_1] \times [y_0, y_1] \times [z_0, z_1]$ uniformly discretised into $N_x$, $N_y$, and $N_z$ grid points along each direction. The corresponding grid spacing is defined as
\begin{equation}
    h_x = \frac{x_1-x_0}{N_x-1}, \quad h_y = \frac{y_1-y_0}{N_y-1}, \quad h_z = \frac{z_1-z_0}{N_z-1}.
\end{equation}

Different quantities, such as electric and magnetic fields or particle velocities, can be stored on the grid. This functionality was extended to include electron, positive ion, and negative ion density maps.

To populate these maps, three dedicated methods (\texttt{AddIon}, \texttt{AddNegativeIon}, and \texttt{AddElectron}) were implemented following Particle-in-Cell (PIC) methods~\cite{dawson_1983, villasenor_1992, birdsall_2018, hockney_2021}. The weighting, or deposition, process consists of distributing the particle contribution onto the surrounding grid nodes using a first-order shape function $S^{(1)}$. The following expressions correspond to the one-dimensional case, where $x$ denotes the particle coordinate and $X_i$ the coordinate of the $i$-th node ($i=0,\ldots,N_x-1$).
\begin{equation}
    S^{(1)}(x-X_i)=
        \begin{cases}
            1 - \abs{\frac{x-X_i}{h_x}}, &\qif \abs*{x-X_i} \le h_x \\
            0, &\qotherwise
        \end{cases}
\end{equation}

Particle density $n_i$ at node $i$ is computed using the shape function $S^{(1)}$, where $N_p$ is the number of simulated particles and $x_k$ and $w_k$ denote the coordinate and weight of particle $k$. The weight represents the multiplicity of the simulated particle, i.e. the number of physical particles it represents.
\begin{equation}
    n_i = \frac{1}{h_x} \sum^{N_p}_k w_k \, S^{(1)}(x_k-X_i)
    \label{eq:particle_density_PIC}
\end{equation}

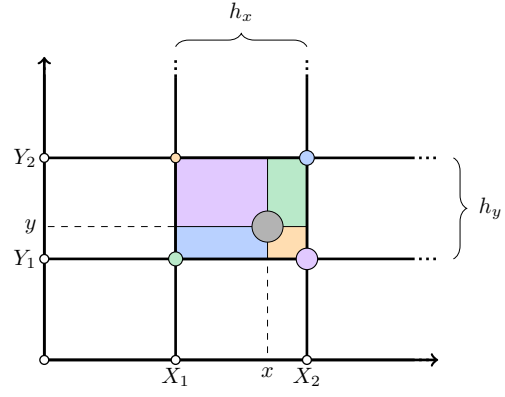
\begin{figure}[t]
	\centering
    \begin{tikzpicture}[x=1.6cm,y=1.6cm]
  \definecolor{myblue}{RGB}{80,140,255}
  \definecolor{mygreen}{RGB}{80,200,120}
  \definecolor{myorange}{RGB}{255,170,60}
  \definecolor{mypurple}{RGB}{180,120,255}

  \def\hx{1.3}
  \def\hy{1.0}

  \def\xm{-1.3}
  \def\xL{0}
  \def\xR{1.3}
  \def\xExt{2.6}
  \def\xDot{2.34}

  \def\ym{-1}
  \def\yB{0}
  \def\yT{1}
  \def\yExt{2}
  \def\yDot{1.8}

  \foreach \x in {\xm,\xL,\xR} {
    \draw[very thick] (\x,\ym) -- (\x,\yDot);
    \draw[very thick,dotted] (\x,\yDot) -- (\x,\yExt);
  }
  \foreach \y in {\ym,\yB,\yT} {
    \draw[very thick] (\xm,\y) -- (\xDot,\y);
    \draw[very thick,dotted] (\xDot,\y) -- (\xExt,\y);
  }

  \draw[->,very thick] (\xm,\ym) -- (\xExt,\ym);
  \draw[->,very thick] (\xm,\ym) -- (\xm,\yExt);

  \foreach \x in {\xm,\xL,\xR}
    \foreach \y in {\ym,\yB,\yT}
      \fill[white,draw=black, line width=0.6pt] (\x,\y) circle (2.pt);

  \def\xSplit{0.91} 
  \def\ySplit{0.32}

  \draw[dashed] (\xSplit,\ySplit) -- (\xSplit,\ym);
  \node[below] at (\xSplit,\ym) {$x$};
  \draw[dashed] (\xSplit,\ySplit) -- (\xm,\ySplit);
  \node[left] at (\xm,\ySplit) {$y$};

  \node[below] at (\xL,\ym) {$X_1$};
  \node[below] at (\xR,\ym) {$X_2$};
  \node[left] at (\xm,\yB) {$Y_1$};
  \node[left] at (\xm,\yT) {$Y_2$};

  \draw[decorate,decoration={brace,amplitude=6pt}]
    (\xL,2.1) -- (\xR,2.1)
    node[midway,above=8pt] {$h_x$};
  \draw[decorate,decoration={brace,mirror,amplitude=6pt}]
      (2.75,\yB) -- (2.75,\yT)
      node[midway,right=8pt] {$h_y$};

  \fill[mypurple!40] (\xL,\ySplit) rectangle (\xSplit,\yT);
  \fill[myblue!40] (\xL,\yB) rectangle (\xSplit,\ySplit);
  \fill[mygreen!40] (\xSplit,\ySplit) rectangle (\xR,\yT);
  \fill[myorange!40] (\xSplit,\yB) rectangle (\xR,\ySplit);

  \draw[very thick] (\xL,\yB) rectangle (\xR,\yT);

  \draw[] (\xSplit,\yB) -- (\xSplit,\yT);
  \draw[] (\xL,\ySplit) -- (\xR,\ySplit);

  \fill[gray!60,draw=black, line width=0.4pt]
      (\xSplit,\ySplit) circle (7pt);

  \fill[myorange!40,draw=black,line width=0.5pt]
      (\xL,\yT) circle (2.17pt); 

  \fill[myblue!40,draw=black,line width=0.5pt]
      (\xR,\yT) circle (3.31pt); 

  \fill[mygreen!40,draw=black,line width=0.5pt]
      (\xL,\yB) circle (3.16pt); 

  \fill[mypurple!40,draw=black,line width=0.5pt]
      (\xR,\yB) circle (4.83pt); 
\end{tikzpicture}
	\caption{Schematic illustration of the Particle-in-Cell (PIC) deposition scheme in a two-dimensional grid.}
	\label{fig:PIC}
\end{figure}

\cref{fig:PIC} provides a visual illustration of the deposition scheme in a two-dimensional grid. The grey particle located inside a grid cell distributes its weight to the four surrounding nodes according to its relative position, as indicated by the coloured regions.

To retrieve the value of a scalar field quantity at a given spatial position, the pre-existing method \texttt{GetData} is used. This method performs a three-dimensional interpolation of a predefined discrete map using a first-order shape function. This interpolation step, which complements the weighting procedure, not only smooths the discrete data but also provides a more accurate and spatially continuous estimation of the field between grid points.

\subsection{Implementation of space charge}
\label{sec:implementation_space_charge}
In order to account for the space-charge effects arising from the high charge densities generated under UHDR conditions, we implemented a Poisson solver within the \texttt{ComponentGrid} class. Similarly to the ion density maps, a \texttt{ComponentGrid} object can now store a space-charge density $\rho$ map computed according to~\cref{eq:charge_density_PIC}, where $q_k$ denotes the charge of particle $k$.
\begin{equation}
    \rho_i = \frac{1}{h_x} \sum^{N_p}_k q_k \, w_k \, S^{(1)}(x_k-X_i)
    \label{eq:charge_density_PIC}
\end{equation}

A new method, \texttt{ComputeField}, was added to this class to solve Poisson's equation and update the corresponding electric field. The electrostatic potential $\phi$ generated by the space-charge density $\rho$ is obtained from
\begin{equation}
    \laplacian{\phi} = - \frac{\rho}{\varepsilon},
    \label{eq:Poisson}
\end{equation}
where $\varepsilon$ denotes the permittivity of the medium.

To discretise the Laplacian, we used a central difference approximation of the second derivative.  In one dimension, this results in the following formula:
\begin{multline}
    f''(x) = \frac{f(x - h_x) - 2f(x) + f(x + h_x)}{h_x^2} \\
    + \order{h_x^2}.
    \label{eq:discretised_Laplacian}
\end{multline}
On a three-dimensional grid, Poisson’s equation can then be written as:
\begin{multline}
     -\frac{\rho_{i,j,k}}{\varepsilon} = \frac{\phi_{i-1,j,k} -2\,\phi_{i,j,k} + \phi_{i+1,j,k}}{h_x^2} \\
     + \frac{\phi_{i,j-1,k} -2\,\phi_{i,j,k} + \phi_{i,j+1,k}}{h_y^2} \\
     + \frac{\phi_{i,j,k-1} -2\,\phi_{i,j,k} + \phi_{i,j,k+1}}{h_z^2}.
    \label{eq:discretised_Poisson}
\end{multline}

To solve Poisson's equation, appropriate boundary conditions must be specified. In this work, we use the general Robin boundary condition defined in \cref{eq:robin_BC}, where $n$ denotes the outward normal direction. The discretised form of this boundary condition is provided in \cref{app:bc}.
\begin{equation}
    \alpha \, \phi + \beta \, \frac{\partial \phi}{\partial n} = \gamma
    \label{eq:robin_BC}
\end{equation}
It is clear from \cref{eq:robin_BC} that the Robin boundary conditions includes Dirichlet and Neumann boundary conditions as particular cases.

Dirichlet boundary conditions, obtained by setting $\alpha = 1$ and $\beta = 0$, fix the value of the electric potential to $\gamma$ at the boundaries of the domain $\Omega$. This type of condition is used to impose the electrode potentials, which are located on the planes normal to the $x$-axis. The cathode, located at $x=x_0$, is grounded, while the anode, located at $x=x_1$, is held at a potential $U$:
\begin{equation}
    \phi(x_0) = 0,
    \qquad
    \phi(x_1) = U.
\end{equation}

Neumann boundary conditions, obtained by setting $\alpha = 0$ and $\beta = 1$, specify the value of the normal derivative of the potential at the boundary:
\begin{equation}
    \frac{\partial \phi}{\partial n} = \gamma.
\end{equation}

In the homogeneous case, this derivative is set to zero, meaning that no electric flux crosses the boundary. Physically, a homogeneous Neumann boundary condition corresponds to an insulating wall, where the normal component of the electric field vanishes. In the case of a chamber with electrodes, this means that no electric flux (and thus no charge) can escape through the side boundaries, and the field lines near the walls remain parallel to them. For the faces normal to the $y$- and $z$-axes, we therefore impose:
\begin{equation}
    \pdv{\phi(y_0,\, y_1)}{y} = 0 \quad \text{and} \quad \pdv{\phi(z_0,\, z_1)}{z} = 0.
\end{equation}

We used an iterative Gauss--Seidel solver~\cite{seidel_1874} accelerated by the Successive Over-Relaxation (SOR) method~\cite{young_1954} to solve the discretised Poisson equation~(\cref{eq:discretised_Poisson}). A detailed description of the algorithm is given in \cref{app:gauss_seidel}.

Solving the Poisson equation with an iterative method is computationally advantageous in comparison, for instance, with the particle-to-particle or node-to-node methods. These methods sum the contribution of each node or particle to obtain the electric field at a certain position $(x, y, z)$. The algorithmic complexity is then of the order of $\order{N^2}$ with $N$ being the number of particles or cells. The Gauss--Seidel iterative method loops only $n$ times over the cells, $n$ being the total number of iterations. The complexity is then $\order{Nn}$. In our case, for a residual tolerance of $\num{1e-6}$ and a number of cells of $N=N_x \times N_y \times N_z= 30 \times 30 \times 30$, the number of iterations required for convergence remains reasonable, typically below 100.

\subsection{Code structure, parallelisation and optimisation}
First, the class \texttt{MediumMagboltz} is used to define the gas mixture by interfacing Magboltz. The composition, pressure, temperature, and ion mobility can be set. This defines the medium in which the simulation takes place. The chamber applied electric field is managed by the \texttt{ComponentAnalyticField} class. The two electrodes are defined and the electric field is calculated analytically for the simple geometry of the PPIC. The electrodes are assumed to be infinitely flat and, consequently, no edge effects are taken into account. The \texttt{ComponentGrid} class is then used to discretise the geometry and compute the space-charge electric field and the particle density maps. Finally, the \texttt{Sensor} class allows the different components to be combined. The total electric field is the sum of the electric fields of each component. This class also allows the calculation of currents induced by drifting charged particles using the Shockley--Ramo theorem~\cite{shockley:1938, ramo:1939}.

The pulse structure can be recreated in two different ways: either by assuming that ionisation is homogeneous and isotropic in the chamber and distributing the electron--ion pairs accordingly, or by using the \texttt{TrackHeed} class to simulate the primary ionisation caused by the passage of charged particles through the detector. Both the temporal and spatial structures of the pulse can easily be modified by changing the pulse duration or the spatial distribution of pairs or incident particles.

In all charge-transport simulations presented here, electron--ion pairs were generated directly and uniformly within the irradiated volume. A square temporal pulse profile was used, except for the comparison with the Boag and Fenwick models in~\cref{sec:Fenwick}, for which the pairs were generated instantaneously. Primary ionisation by incident radiation was not simulated with Heed. Consequently, no incident particle type or energy was explicitly assigned. Irradiation conditions were represented through the number of generated pairs and their spatial and temporal distributions, with the corresponding absorbed dose calculated as described in~\cref{app:dose}.

To reduce the computational cost associated with the large number of electron--ion pairs generated under the investigated irradiation conditions, a multiplicity factor, referred to as the particle weight and introduced in \cref{sec:implementation_recombination}, was used. Each simulated electron or ion therefore represents multiple physical particles.

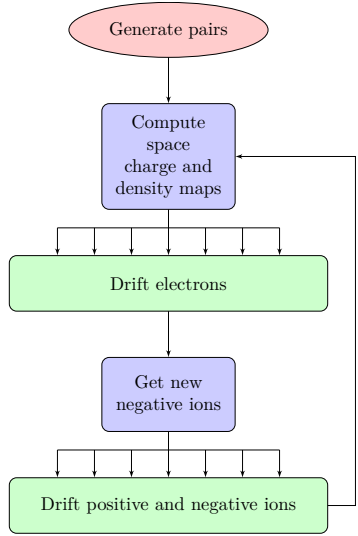
\begin{figure}[t]
	\centering
    \tikzstyle{block} = [rectangle, draw, fill=blue!20,
  text width=6em, text centered, rounded corners,
  minimum height=4em, inner sep=6pt, outer sep=0pt]

\tikzstyle{largeblock} = [rectangle, draw, fill=green!20,
  text width=16em, text centered, rounded corners,
  minimum height=3em, inner sep=6pt, outer sep=0pt]

\tikzstyle{cloud} = [draw, ellipse, fill=red!20,
  minimum height=2em, inner sep=6pt, outer sep=0pt, node distance = 2em]

\tikzstyle{line} = [draw, -latex']
\tikzstyle{decision} = [diamond, draw, fill=blue!20, 
    text width=4.5em, text badly centered, node distance=4cm, inner sep=6pt, outer sep=0pt]

\begin{tikzpicture}[node distance = 2.5em, auto]
    \node [cloud] (input) {Generate pairs};
    \node [block, below= of input] (computeField) {Compute space charge and density maps};
    \node [largeblock, below= of computeField] (drifte) {Drift electrons};
    \node [block, below= of drifte] (inputnegions) {Get new negative ions};
    \node [largeblock, below= of inputnegions] (drifti) {Drift positive and negative ions};
 
    \path [line] (input) -- (computeField);

    \path [line] (computeField) -- (drifte);
    \draw [-] ([xshift=-6em, yshift=-1em]computeField.south) -- ([xshift=6em, yshift=-1em]computeField.south);
    \path [line] ([xshift=2em, yshift=-1em]computeField.south) -- ([xshift=2em]drifte.north);
    \path [line] ([xshift=-2em, yshift=-1em]computeField.south) -- ([xshift=-2em]drifte.north);
    \path [line] ([xshift=4em, yshift=-1em]computeField.south) -- ([xshift=4em]drifte.north);
    \path [line] ([xshift=-4em, yshift=-1em]computeField.south) -- ([xshift=-4em]drifte.north);
    \path [line] ([xshift=6em, yshift=-1em]computeField.south) -- ([xshift=6em]drifte.north);
    \path [line] ([xshift=-6em, yshift=-1em]computeField.south) -- ([xshift=-6em]drifte.north);

    \path [line] (drifte) -- (inputnegions);

    \path [line] (inputnegions) -- (drifti);
    \draw [-] ([xshift=-6em, yshift=-1em]inputnegions.south) -- ([xshift=6em, yshift=-1em]inputnegions.south);
    \path [line] ([xshift=2em, yshift=-1em]inputnegions.south) -- ([xshift=2em]drifti.north);
    \path [line] ([xshift=-2em, yshift=-1em]inputnegions.south) -- ([xshift=-2em]drifti.north);
    \path [line] ([xshift=4em, yshift=-1em]inputnegions.south) -- ([xshift=4em]drifti.north);
    \path [line] ([xshift=-4em, yshift=-1em]inputnegions.south) -- ([xshift=-4em]drifti.north);
    \path [line] ([xshift=6em, yshift=-1em]inputnegions.south) -- ([xshift=6em]drifti.north);
    \path [line] ([xshift=-6em, yshift=-1em]inputnegions.south) -- ([xshift=-6em]drifti.north);

    \draw[line] (drifti.0) -- ++(1.5em, 0) |- (computeField);
\end{tikzpicture}
	\caption{Diagram of the main components of the simulation code. Multithreaded sequences are shown in green, while single-threaded sequences are shown in blue and red.}
	\label{fig:parallel_diagram}
\end{figure}

Garfield++ is designed such that each particle drift is simulated independently and sequentially. While this structure is advantageous for parallelisation, as used in this work, it becomes problematic when computing particle and charge density maps, which require coordinated access to shared data structures, namely the particle positions. To overcome this issue, we defined a global time step after which the parallelised particle drifts are stopped, allowing the computation of the particle and charge density maps. The corresponding recombination probabilities and the space-charge electric field derived from these maps are then used throughout the entire time step. \cref{fig:parallel_diagram} shows a diagram of the different sequences of the code.

\section{Results and discussion}
\subsection{Charge collection efficiency without space charge}
\label{sec:Fenwick}

In \citeyear{fenwick_2023}, \textcite{fenwick_2023} revisited Boag’s model proposed in \citeyear{boag_1996} and highlighted one of its main limitations: the oversimplified assumption regarding the spatial distribution of anions.

To derive an analytical expression for the charge collection efficiency (CCE), \textcite{boag_1996} replaced the physical anion distribution described by \cref{eq:anion_distribution} with a simplified model assuming a uniform anion concentration within a given region and zero concentration elsewhere. Modern numerical methods now make it possible to relax this assumption and use a more realistic spatial distribution of negative ions for instantaneous pulses while neglecting space-charge effects.

\begin{equation}
    \frac{\dd n_{-}}{\dd x} = \eta \, n_\text{e}
	\label{eq:anion_distribution}
\end{equation}

\cref{eq:anion_distribution} describes the spatial evolution of the anion density $n_{-}$, governed by the electron attachment coefficient $\eta$ in \si{\per\m} and electron density $n_\text{e}$.

The first validation of our simulation framework was performed against the model proposed by \textcite{fenwick_2023}. Since this model does not include space-charge effects, these were disabled in the Monte Carlo simulations. 

\begin{figure}[t]
	\centering
    \includegraphics[width=\linewidth]{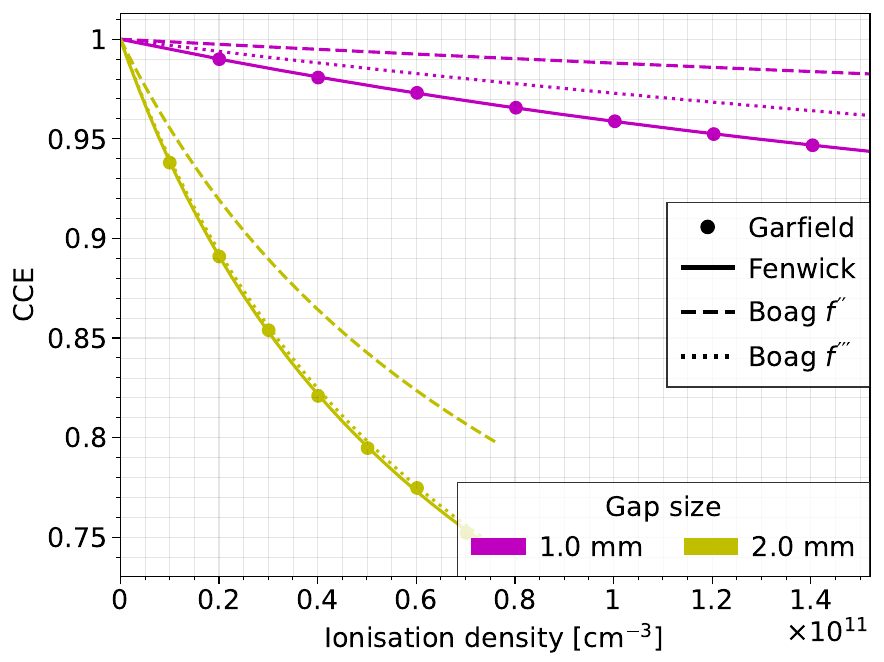}
	\caption{Charge collection efficiency as a function of ionisation density for an instantaneous pulse in PPICs with \SI{1}{\mm} and \SI{2}{\mm} gaps operated at \SI{500}{\V}. Garfield++ results with space-charge effects disabled (markers) are compared with the Fenwick and Boag models (lines).}
	\label{fig:results_fenwick}
\end{figure}

\Cref{fig:results_fenwick} compares the present Monte Carlo simulation with the exact solution proposed by \textcite{fenwick_2023} and with Boag's models ($f''$ and $f'''$) for two electrode gap sizes. Each data point corresponds to the mean value of 40 independent Monte Carlo simulations. Simulations were performed for PPICs with electrode gaps of \SI{1.0}{\mm} and \SI{2.0}{\mm}, both operated at an applied voltage of \SI{500}{\volt}. The grid was discretised into 25 cells along each spatial direction and updated every \SI{10}{\nano\second}.

In each simulation, \num{50000} electron--ion pairs were generated in an instantaneous pulse, with particle weights ranging linearly from \numrange{1260}{8820} to reproduce the targeted ionisation densities. The beam was modelled as spatially uniform with a circular cross section of radius \SI{1.0}{\mm}. Using parallel execution on 10 CPUs, the average runtime of a single simulation was approximately \SI{7}{\minute} and \SI{25}{\minute} for the \SI{1.0}{\mm} and \SI{2.0}{\mm} gap configurations, respectively.

The Fenwick model, which depends on the attachment coefficient, was then fitted to the simulation results. The relative difference between the FEF obtained from the Monte Carlo simulation and that derived from the fitted Fenwick model was found to be \SI{0.03}{\percent} for the \SI{1}{\mm} gap and \SI{0.20}{\percent} for the \SI{2}{\mm} gap, demonstrating excellent agreement between the two approaches.

\subsection{Space charge effect}
The Poisson solver implemented in \cref{sec:implementation_space_charge} was first validated against an analytical test case. We considered a uniform distribution of \num{1e7} ions confined within a cylindrical volume aligned along the $x$-axis, with radius \SI{1.0}{\mm} and length \SI{1.6}{\mm}. This configuration provides a known analytical solution for the electric field while remaining representative of charge distributions encountered in PPICs.

To approximate open boundary conditions within a finite computational domain, mixed boundary conditions were applied. Following~\cref{eq:robin_BC}, we fixed $\gamma = 0$ and $\beta = 1$.
\begin{equation}
    \alpha \, \phi + \frac{\partial \phi}{\partial n} = 0
\end{equation}

This first-order asymptotic boundary condition~\cite{meeker_2014, jin_2014} approximates a vanishing potential at infinity. The asymptotic formulation leads to a Robin coefficient scaling with the inverse of a characteristic domain length. Accordingly, we used $\alpha=C/L_x$, where $L_x$ denotes the size of the computational domain along the $x$-direction. The dimensionless coefficient $C=1.4$ was determined empirically by minimising the discrepancy between the numerical electric field and the analytical solution for the uniformly charged cylinder used in this validation test.

\begin{figure}[t]
	\centering
    \includegraphics[width=\linewidth]{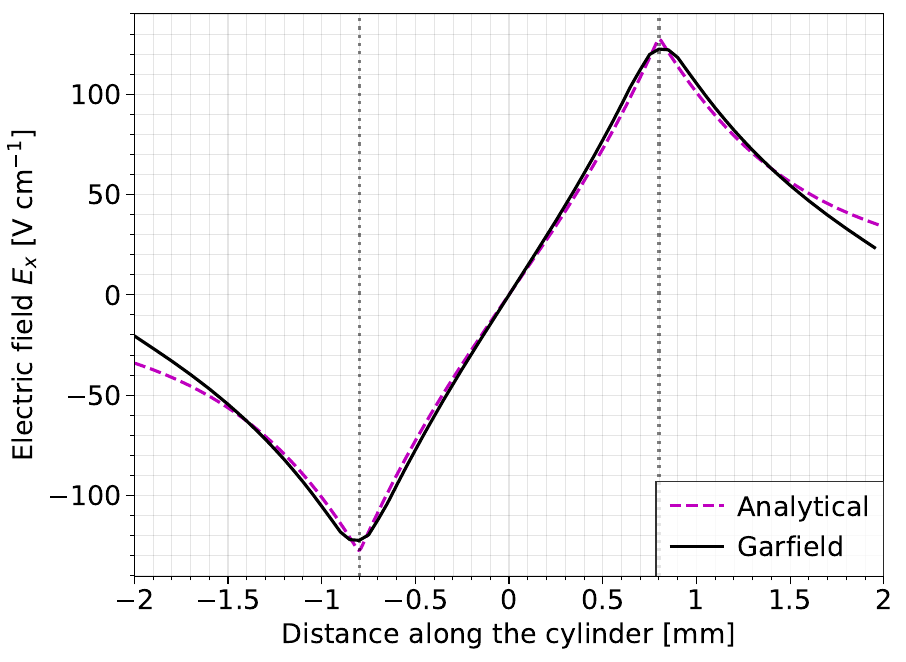}
	\caption{Comparison between the electric field computed using the Poisson solver (red) and the corresponding analytical solution (black) for a uniformly charged cylindrical distribution. The dashed lines are the limit of the cylinder.}
	\label{fig:field_solver_comparison}
\end{figure}

The electric field obtained from the numerical solution is compared with the analytical result in \cref{fig:field_solver_comparison}. Excellent agreement is observed throughout most of the domain, with only minor deviations near the boundaries attributable to the finite discretisation and the approximate nature of the boundary conditions.

Following the validation of the solver, we investigated the temporal evolution of the electric field inside the chamber after a square, spatially uniform pulse of \SI{2.5}{\micro\second} delivering a dose of \SI{2}{\gray}. The chamber consisted of a \SI{1}{\mm} gap operated at a bias voltage of \SI{300}{\volt}.

\begin{figure*}[t]
\centering
\begin{subfigure}[t]{0.48\textwidth}
    \centering
    \includegraphics[width=\linewidth]{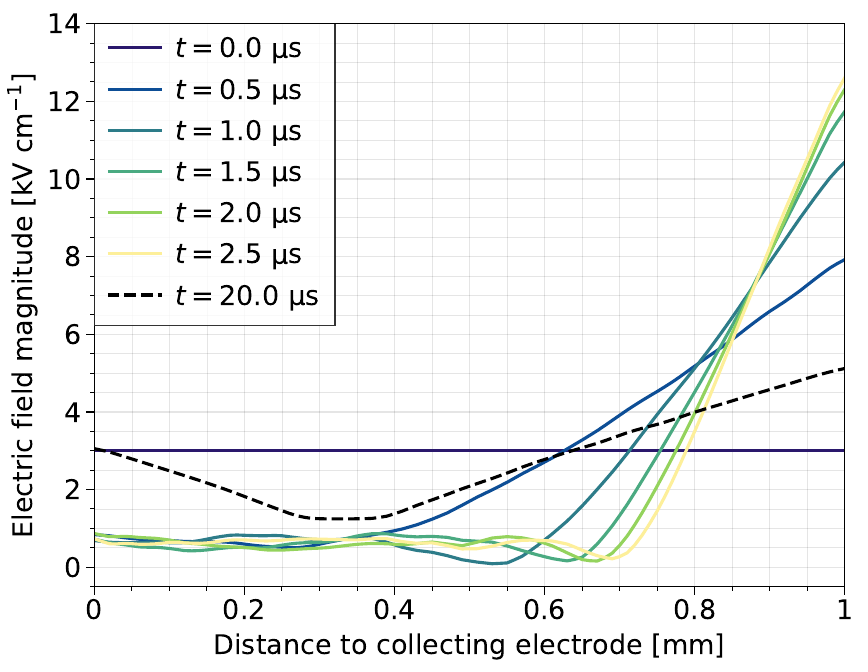}
    \caption{Garfield++ simulation for a spatially uniform pulse with a square temporal profile} 
    \label{fig:scs_time_evolution_garfield}
\end{subfigure}
\hfill
\begin{subfigure}[t]{0.48\textwidth}
    \centering
    \includegraphics[width=\linewidth]{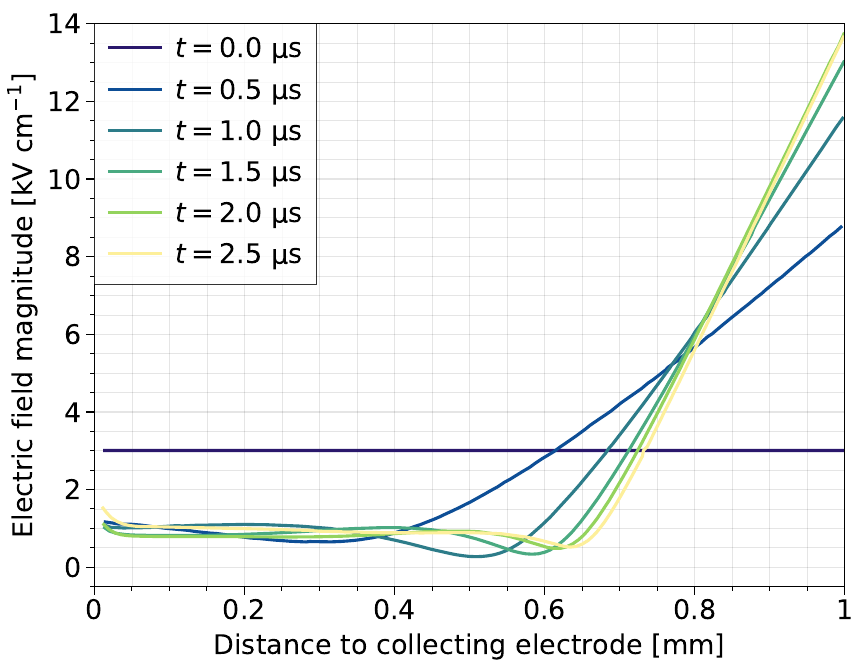}
    \caption{One-dimensional simulation from \textcite{kranzer_2021}.}
    \label{fig:scs_time_evolution_kranzer}
\end{subfigure}
\caption{Temporal evolution of the electric field magnitude along the beam isocentre in a \SI{1}{\milli\meter} gap PPIC operated at \SI{300}{\volt}, for a pulse delivering \SI{2}{\gray} over \SI{2.5}{\micro\second}.}
\label{fig:scs_time_evolution}
\end{figure*}

\Cref{fig:scs_time_evolution_garfield} shows the temporal evolution of the electric field magnitude along the beam isocentre. The results demonstrate that the electric field can be strongly distorted by space charge, not only during the irradiation pulse but also throughout the subsequent charge drift phase. In particular, the electric field can locally exceed four times the nominal applied field, while dropping to nearly zero in other regions of the chamber.

Moreover, very good agreement is observed with the results reported by \textcite{kranzer_2021}. As illustrated in \cref{fig:scs_time_evolution}, the temporal evolution of the electric field is nearly identical in both simulations for the same chamber geometry and pulse conditions.

These results confirm that, under UHDR irradiation conditions, space-charge effects can significantly distort the electric field inside ionisation chambers. Such perturbations strongly affect electron and ion transport properties, including drift velocity, mobility, and electron attachment coefficient, thereby modifying the free electron fraction.

\subsection{Free electron fraction}

Following the work of \textcite{paz-martin_2022}, we investigated the evolution of the FEF under UHDR irradiation conditions. In Boag’s theory, the FEF is treated as a constant parameter characteristic of a given IC operated at a fixed bias voltage. However, strong space-charge-induced distortions of the electric field can invalidate this assumption at high dose per pulse (DPP).

For a chamber with an electrode gap $d$, the FEF $p$ is given by
\begin{equation}
	p = \frac{1}{\eta \, d}\left(1 - \exp(-\eta \, d)\right)
	\label{eq:fef}
\end{equation}
where $\eta$ denotes the electron attachment coefficient.

Under UHDR conditions, the electric field inside the chamber is no longer spatially uniform. Consequently, the attachment coefficient $\eta$, which strongly depends on the electric field strength as shown in \cref{fig:attachment_and_mobility}, also becomes spatially and temporally dependent.

\begin{figure}[t]
	\centering
    \includegraphics[width=\linewidth]{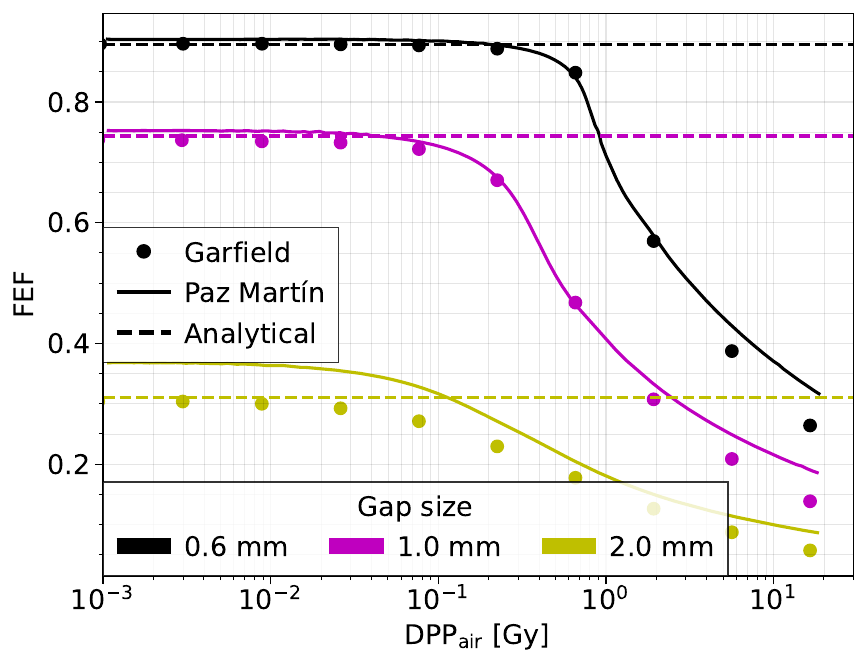}
	\caption{Free electron fraction as a function of DPP for different gap sizes of PPICs at \SI{300}{\V} and a pulse duration of \SI{2.5}{\micro\second}. Results from the present simulation are compared with those reported by \textcite{paz-martin_2022} and the constant analytical model.}
	\label{fig:results_paz-martin_FEF}
\end{figure}

\Cref{fig:results_paz-martin_FEF} shows the FEF as a function of DPP for different electrode gap sizes operated at \SI{300}{\volt} and irradiated with a \SI{2.5}{\micro\second} pulse (the conversion from the number of electron--ion pairs to absorbed dose in air is detailed in~\cref{app:dose}). At low DPP, the FEF remains constant and agrees with the analytical prediction given by \cref{eq:fef}. However, as the DPP increases, the FEF progressively decreases.

This behaviour can be explained by the strong electric field distortion induced by space charge. As shown in \cref{fig:scs_time_evolution}, the electric field is reduced in large regions of the chamber during irradiation. Since lower electric field strengths correspond to larger attachment coefficients, the electron attachment probability increases, thereby reducing the fraction of free electrons reaching the electrodes.

The results reported by \textcite{paz-martin_2022} are also shown in \cref{fig:results_paz-martin_FEF}. Excellent agreement is observed at low DPP, while the decrease of the FEF at higher DPP follows the same trend in both simulations. The remaining discrepancy observed for the \SI{2.0}{\mm} gap configuration, even at low DPP, is likely attributable to differences in the attachment coefficient used in the simulations.

\subsection{Charge collection efficiency}
To further validate our simulation framework under UHDR conditions, we investigated the CCE in the presence of space-charge effects and compared our results with those reported by \textcite{paz-martin_2022}. In their work, the authors validated their numerical model against experimental measurements obtained for several chamber bias voltages, providing a reliable reference for comparison.

The CCE was computed as a function of DPP for a PPIC with a \SI{2}{\mm} electrode gap irradiated with a \SI{2.5}{\micro\second} pulse.

\begin{figure}[t]
	\centering
    \includegraphics[width=\linewidth]{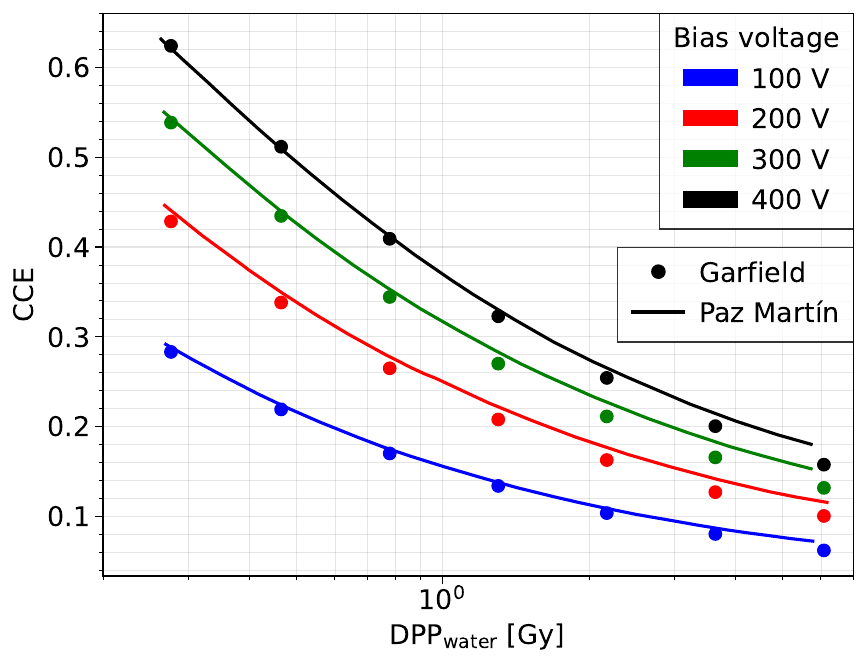}
	\caption{Charge collection efficiency as a function of DPP for a PPIC with a \SI{2}{\mm} electrode gap at four applied voltages and a pulse duration of \SI{2.5}{\micro\second}. Results from the present simulation are compared with those reported by \textcite{paz-martin_2022}.}
	\label{fig:results_paz-martin_CCE}
\end{figure}

\Cref{fig:results_paz-martin_CCE} compares the CCE obtained from the present Monte Carlo simulation with the reference results for four applied voltages. Very good agreement is observed over the full investigated DPP range, demonstrating the ability of the model to accurately reproduce recombination effects in the presence of strong space-charge-induced electric-field distortions under UHDR irradiation conditions.

\subsection{Induced current}
\begin{figure}[t]
	\centering
    \includegraphics[width=\linewidth]{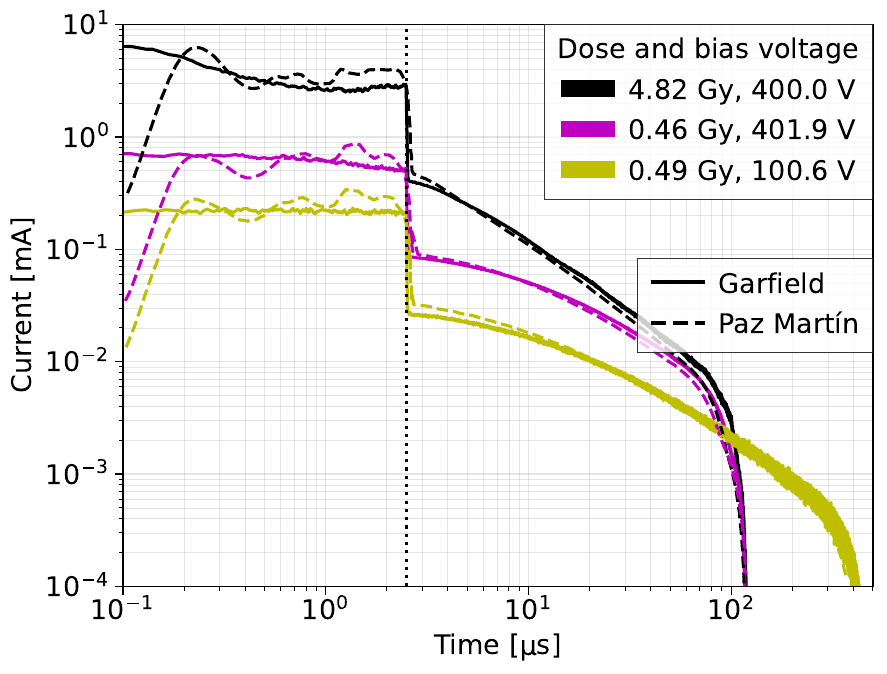}
	\caption{Induced current generated by the drift of electrons and ions. Results from the present simulation compared with those reported by \textcite{paz-martin_2022}. For a \SI{2.5}{\micro\second} pulse in a PPIC with a \SI{2}{\milli\meter} electrode gap and a radius of \SI{7.8}{\mm}.}
	\label{fig:results_paz-martin_current}
\end{figure}

Finally, the induced current generated by the drift of electrons and ions was computed using the Shockley--Ramo theorem. \Cref{fig:results_paz-martin_current} shows the fast electron signal up to the end of the irradiation pulse (black dashed line), followed by the long ion-induced current tail. The simulation was performed for a spatially uniform square pulse of \SI{2.5}{\micro\second} in a PPIC with a \SI{2}{\milli\meter} electrode gap and a radius of \SI{7.8}{\mm}.

The ion-induced current shows very good agreement with the results reported by \textcite{paz-martin_2022}. Small discrepancies are nevertheless observed for the electron signal, particularly during the rising edge at the beginning of the pulse. These differences are most likely attributable to the temporal profile of the irradiation pulse. In the reference study, the experimentally measured temporal pulse shape was used as an input to the numerical simulation, whereas a square temporal profile is assumed in the present work. This difference mainly affects the fast electron-induced component of the signal, while having a much smaller influence on the subsequent ion-induced current.

This result also confirms a surprising observation reported in the reference study: the DPP does not significantly affect the charge collection time. This indicates that the reduction of the CCE under UHDR conditions cannot be explained by a longer coexistence time of positive and negative ions within the chamber, despite the strong space-charge-induced electric-field distortions. Instead, the reduction of the CCE appears to be primarily driven by the decrease of the FEF caused by the space-charge-induced modification of the attachment coefficient. This mechanism increases the density of negative ions and consequently enhances ion--ion recombination probabilities.

\subsection{Uncertainties}

As discussed in \cref{sec:ion_swarm_parameters}, the dominant physical uncertainties in the simulation arise from the ion mobility and the ion--ion recombination coefficient, since these parameters directly govern ion transport and recombination rates within the chamber.

Additional uncertainties originate from the numerical implementation of the Monte Carlo simulation. To maintain a reasonable computational cost, weighted particles are employed, with each simulated particle representing a large number of physical ions or electrons. This approximation introduces statistical fluctuations associated with the finite number of simulated particles.

Further uncertainties arise from the representation of particles and charge densities on a discrete voxel grid. Since local charge densities, recombination probabilities, and space-charge electric fields are evaluated on this grid, the simulation results depend on the chosen spatial resolution. A finer voxelisation improves the spatial description of the charge distribution but simultaneously increases statistical fluctuations due to the reduced number of particles contained in each voxel.

Additional numerical uncertainties are associated with simulation parameters such as the particle transport time step and the update frequencies of grid, particle density maps, and space-charge electric field. These parameters affect both the numerical accuracy and the computational cost of the simulation.

A comprehensive uncertainty analysis covering all simulation parameters would require a prohibitively large number of simulations and is therefore beyond the scope of the present work. Nevertheless, a limited sensitivity study performed for a representative irradiation configuration showed maximum observed variations of \SI{1.24}{\percent} in the FEF and \SI{0.82}{\percent} in the CCE across the grid resolutions tested. These results characterise the sensitivity to spatial discretisation within the tested configuration and do not constitute a comprehensive uncertainty estimate for the final CCE or the inferred dose.
\section{Conclusion}

In this work, the two main physical mechanisms missing from Garfield++ for the simulation of PPICs under UHDR conditions were implemented: ion–ion recombination and space-charge-induced electric-field distortions. Under FLASH irradiation conditions, the large charge densities produced during the pulse strongly perturb the electric field inside the chamber, thereby affecting the CCE. As a consequence, the standard analytical recombination models commonly used in conventional radiotherapy are no longer valid in the UHDR regime.

The model proposed by \textcite{fenwick_2023}, which provides an exact solution for the CCE of an instantaneous ionisation pulse in the absence of space-charge effects, was successfully reproduced. Although this model improves upon Boag’s formalism, its applicability remains limited to instantaneous pulses and low-DPP conditions for which space-charge effects are negligible.

The simulations reported by \textcite{kranzer_2021}, describing the temporal evolution of space-charge-induced electric-field distortions inside ionisation chambers, were also reproduced. For a \SI{2}{\gray} pulse delivered over \SI{2.5}{\micro\s} in a chamber with a \SI{1}{\mm} electrode gap, the electric field magnitude can locally increase by more than a factor of four or decrease to nearly zero in some regions of the detector. Moreover, these distortions remain significant even after complete electron collection.

The results reported by \textcite{paz-martin_2022} regarding the evolution of the FEF, the CCE as a function of DPP, and the induced current were also successfully reproduced. Beyond this validation, our simulations provide evidence that the reduction of the CCE under UHDR conditions is primarily governed by the evolution of the FEF.

The results suggest that this behaviour originates from the strong dependence of the electron attachment coefficient on the local electric field. Space-charge-induced electric-field distortions reduce the field magnitude in large regions of the chamber, thereby increasing electron attachment probabilities. This leads to a lower FEF, a higher density of negative ions, and consequently enhanced ion–ion recombination. In contrast, the simulations indicate that the reduction of the CCE cannot be explained by a longer coexistence time of positive and negative ions within the chamber.

This observation suggests that the UHDR dosimetry problem in ionisation chambers may be largely reduced to the accurate description of the FEF evolution. If confirmed by future studies, this could open the way to improved analytical models or corrections to existing recombination theories~\cite{boag_1996, fenwick_2023}, enabling simple and potentially real-time corrections of the CCE in PPICs under UHDR irradiation conditions.

The implementation of these physical mechanisms within Garfield++ provides a complete, flexible, and self-consistent Monte Carlo framework capable of accurately reproducing UHDR effects in ionisation chambers. The flexibility of Garfield++ enables simulations for arbitrary gas mixtures, atmospheric conditions, detector geometries, incident particle types, and beam spatial and temporal structures. In addition, the primary ionisation produced by different incident particles can be simulated using Heed~\cite{smirnov_2005}.

In contrast to the previous numerical approaches of~\textcite{kranzer_2021} and \textcite{paz-martin_2022} which are based on the solution of one-dimensional charge-carrier transport equations, the present framework follows charge transport using a three-dimensional particle-based Monte Carlo approach. Charge-carrier densities and the resulting space-charge electric field are reconstructed self-consistently from the simulated particle distributions, without assuming transverse uniformity. This approach provides greater flexibility for investigating the influence of the spatial and temporal beam structure and, through the integration of Heed and Magboltz within Garfield++, allows different incident particle types, gas compositions, atmospheric conditions, detector geometries, and irradiation conditions to be considered within the same framework.

The framework developed in this work therefore provides a robust tool for investigating recombination phenomena in ionisation chambers and constitutes a promising basis for the development of next-generation correction models for FLASH dosimetry.

\section*{Code availability}

The two main physical processes implemented in this work, namely ion--ion recombination and space-charge electric-field calculations based on spatial grid discretisation, have been integrated into the open-source Garfield++ software (\url{https://garfieldpp.docs.cern.ch}). These features are available starting from the June 16, 2026 release of Garfield++. An example and tutorial illustrating their use have also been added to the official documentation (\url{https://garfieldpp.docs.cern.ch/tutorials/ionizationchamber/}).

The complete simulation framework developed in this work, named GarFLIC (Garfield++ for FLASH Ionisation Chamber), is openly available at \url{https://gitlab.com/blond-bay/garflic/tree/1.0.0}.

\section*{Acknowledgements}
This work and the associated doctoral research of Pierre Gérard Ortega were funded by the Walloon Region through the PIT ProtherWal programme under grant agreement No. 7289.

G. De Lentdecker is a Research Director of the Fonds de la Recherche Scientifique - FNRS.

Computational resources have been provided by the Consortium des Équipements de Calcul Intensif (CÉCI), funded by the Fonds de la Recherche Scientifique de Belgique (F.R.S.-FNRS) under Grant No. 2.5020.11 and by the Walloon Region.

The authors gratefully acknowledge José Paz Martín for his valuable support and insightful discussions. He kindly provided detailed information on the simulation parameters used in his work and generously answered numerous questions, which greatly contributed to this study.

\printbibliography[title={References}, heading=bibintoc]

\clearpage
\onecolumn
\appendix
\section{Discretisation of Robin boundary conditions}
\label{app:bc}
For simplicity, only the one-dimensional case is presented here. The extension to three dimensions is obtained by applying the same discretisation procedure independently along each coordinate direction.

The Robin boundary condition is defined as
\begin{equation}
    \alpha\,\phi + \beta\,\pdv{\phi}{n} = \gamma,
\end{equation}
where $\vb n$ denotes the outward unit normal vector.

For the left boundary ($\vb n = -\vb{\hat{x}}$), using a first-order forward finite-difference approximation,
\begin{equation}
    \alpha\,\phi_0
    - \beta\,\frac{\phi_1-\phi_0}{h_x}
    = \gamma,
\end{equation}
leading to
\begin{equation}
    \phi_0 = \frac{\gamma h_x + \beta \phi_1}{\alpha h_x + \beta}.
\end{equation}

For the right boundary ($\vb n = \vb{\hat{x}}$), using a first-order backward finite-difference approximation,
\begin{equation}
    \alpha\,\phi_{N_x-1}
    + \beta\,\frac{\phi_{N_x-1}-\phi_{N_x-2}}{h_x}
    = \gamma,
\end{equation}
leading to
\begin{equation}
    \phi_{N_x-1} = \frac{\gamma h_x + \beta \phi_{N_x-2}}{\alpha h_x + \beta}.
\end{equation}

\section{Gauss--Seidel algorithm and Successive Over-Relaxation method}
\label{app:gauss_seidel}
The discretised Poisson equation is solved using a Gauss--Seidel iterative scheme accelerated by the SOR method. For completeness, the main equations underlying the implementation are summarised below.

Given a linear system
\begin{equation}
    \mathcal{A}\vb{x} = \vb{b},
\end{equation}
for a symmetric positive-definite matrix $\mathcal{A}$, the Gauss--Seidel iteration converges regardless of the initial guess $\vb{x}^{(0)}$.

The discrete one-dimensional Laplacian $\laplacian_{N_x} \in \mathbb{R}^{N_x \times N_x}$ can be written as
\begin{equation}
\laplacian_{N_x} = \frac{1}{h_x^2} 
\begin{bmatrix}
-2 & 1      &        &        &   \\
1  & -2     & 1      &        &   \\
   & \ddots & \ddots & \ddots &   \\
   &        & 1      & -2     & 1 \\
   &        &        & 1      & -2
\end{bmatrix},
\end{equation}
and the corresponding three-dimensional discrete Laplacian can then be written as
\begin{equation}
    \laplacian_{N_x,N_y,N_z} =
    I_{N_z} \otimes I_{N_y} \otimes \laplacian_{N_x}
    + I_{N_z} \otimes \laplacian_{N_y} \otimes I_{N_x}
    + \laplacian_{N_z} \otimes I_{N_y} \otimes I_{N_x}.
\end{equation}

Since the discrete Laplacian is symmetric and negative definite, the system is rewritten as
\begin{equation}
    (-\mathcal{A})\vb{x} = -\vb{b},
\end{equation}
yielding a symmetric positive-definite matrix for which Gauss--Seidel convergence is guaranteed.

Given a tolerance $\varepsilon$, convergence is reached when
\begin{equation}
    \frac{\norm*{\vb{r}^{(k)}}}{\norm{\vb{b}}} < \varepsilon,
\end{equation}
with the residual at the $k$-th iteration defined as
\begin{equation}
    \vb{r}^{(k)} = \mathcal{A}\vb{x}^{(k)} - \vb{b}.
\end{equation}

The infinity norm is used as the convergence criterion:
\begin{equation}
    \norm*{\vb{r}^{(k)}}_\infty
    =
    \max_{1 \le i \le n} \abs{r_i^{(k)}},
\end{equation}

From \cref{eq:discretised_Poisson}, we define $A = 1/h_x^2$, $B = 1/h_y^2$, $C = 1/h_z^2$, and $K = -2(A + B + C)$, which allows us to rewrite the discrete Poisson equation as
\begin{equation}
\begin{aligned}
    -\frac{\rho_{i,j,k}}{\varepsilon}
    &=
    A(\phi_{i-1,j,k} + \phi_{i+1,j,k})
    + B(\phi_{i,j-1,k} + \phi_{i,j+1,k}) \\
    &\quad
    + C(\phi_{i,j,k-1} + \phi_{i,j,k+1})
    + K\phi_{i,j,k}.
\end{aligned}
\end{equation}

For convenience, we define
\begin{equation}
\begin{aligned}
    \tau_{i,j,k}
    &=
    A(\phi_{i-1,j,k} + \phi_{i+1,j,k})
    + B(\phi_{i,j-1,k} + \phi_{i,j+1,k}) \\
    &\quad
    + C(\phi_{i,j,k-1} + \phi_{i,j,k+1}),
\end{aligned}
\end{equation}
so that the potential update equation takes the compact form
\begin{equation}
    \phi_{i,j,k}
    =
    \frac{1}{K}
    \left(
    \frac{-\rho_{i,j,k}}{\varepsilon}
    -
    \tau_{i,j,k}
    \right).
\end{equation}

The Gauss--Seidel iteration is then
\begin{equation}
    \tilde{\phi}_{i,j,k}^{(m+1)}
    =
    \frac{1}{K}
    \left(
    \frac{-\rho_{i,j,k}}{\varepsilon}
    -
    \tau_{i,j,k}^{(m+1)}
    \right),
    \label{eq:GS_update_phi}
\end{equation}
where $\tau_{i,j,k}^{(m+1)}$ is given by
\begin{equation}
\begin{aligned}
    \tau_{i,j,k}^{(m+1)}
    &=
    A\left(\phi_{i-1,j,k}^{(m+1)} + \phi_{i+1,j,k}^{(m)}\right)
    + B\left(\phi_{i,j-1,k}^{(m+1)} + \phi_{i,j+1,k}^{(m)}\right) \\
    &\quad
    + C\left(\phi_{i,j,k-1}^{(m+1)} + \phi_{i,j,k+1}^{(m)}\right),
\end{aligned}
\end{equation}
when sweeping the grid points $(i,j,k)$ in lexicographic order.

To further accelerate convergence, we employed the SOR method proposed by \textcite{young_1954}. The updated potential is obtained by adding a fraction of the Gauss--Seidel correction controlled by the relaxation parameter $\omega$:
\begin{equation}
    \phi_{i,j,k}^{(m+1)}
    =
    \phi_{i,j,k}^{(m)}
    +
    \omega
    \left(
    \tilde{\phi}_{i,j,k}^{(m+1)}
    -
    \phi_{i,j,k}^{(m)}
    \right).
\end{equation}

For the configurations considered in this work, SOR reduced the number of iterations required to reach convergence by approximately one order of magnitude compared with the standard Gauss--Seidel scheme.

\section{Absorbed dose}
\label{app:dose}
The absorbed dose $D_\mathrm{m}$ is defined as the energy absorbed $E_\text{ab}$ per unit mass of absorbing medium $\mathrm{m}$. Its SI unit is the gray (\si{\gray}), defined as one joule per kilogram (\si{\joule\per\kilogram}):
\begin{equation}
    D_\mathrm{m} = \frac{\Delta E_\text{ab}}{\Delta m_\mathrm{m}}.
\end{equation}

The absorbed dose in air can be related to the total number of electron--ion pairs $N_0$ created during the ionisation pulse. Let $w$ denote the mean energy required to produce one electron--ion pair in air. The absorbed dose in air is then given by
\begin{equation}
    D_\mathrm{air} = \frac{N_0\,w}{\rho_\mathrm{air}\,\sigma\,d},
\end{equation}
where $\rho_\mathrm{air}$ is the air density, $d$ is the chamber gap, and $\sigma$ is the irradiated area, corresponding either to the beam cross-sectional area or to the chamber sensitive area, whichever is smaller. For the air mixture considered in this work, $w=\SI{33.83}{\eV}$~\cite{magboltz}.

To convert the absorbed dose in air into the absorbed dose in another medium, Bragg--Gray cavity theory is assumed. Under this approximation, the absorbed dose is related to the dose in air through the mass stopping-power ratio $s_{\mathrm{m,air}} = S_\mathrm{m}/S_\mathrm{air}$:
\begin{equation}
    D_\mathrm{m} = s_\mathrm{m,air}\,D_\mathrm{air}.
\end{equation}

\end{document}